# An Innovative Approach for online Meta Search Engine Optimization

Jai Manral and Mohammed Alamgir Hossain

Computational Intelligence Research Group
School of Computing Engineering and Information Sciences
Northumbria University, Newcastle Upon Tyne, NE1 8ST, UK
{jai.manral; alamgir.hossain}@northumbria.ac.uk

This paper presents an approach to identify efficient techniques used in Web Search Engine Optimization (SEO). Understanding SEO factors which can influence page's ranking in search engine is significant for webmasters who wish to attract large number of users to their website. Different from previous relevant research, in this study we developed an intelligent Meta search engine which aggregates results from various search engines and ranks them based on several important SEO parameters. The research tries to establish that using more SEO parameters in ranking algorithms helps in retrieving better search results thus increasing user satisfaction. Initial results generated from Meta search engine outperformed existing search engines in terms of better retrieved search results with high precision.

*Index Terms*— **Search Engine Optimization, Meta Search Engine, Page Ranking, WWW.**

## I. INTRODUCTION

The information on the web is prodigious; searching relevant information is difficult for users. The accuracy of search results is measured by relevancy of query term to web-pages ranked and displayed by search engines. Even a small query term generates thousands of pages, which if not sorted can be difficult to retrieve relevant information. Search engines work hard on structuring and ranking these search results. Different methods and approaches for retrieving and ranking these search results have been suggested. There are several parameters on which search engine assign page ranking to webpages. Search engine optimization is that technique by which web masters try to influence the ranking of search results. There are various factors and parameters for SEO, which can be useful for generating higher rank but too much of optimization, can have adverse effects. SEO can be either White Hat method (suggested by search engines, whose implement improves site content) or Black Hat method (affect page ranking without improving site quality). The focus of this paper is on White Hat SEO techniques, which will be plainly called SEO throughout the paper.

World Wide Web is an ocean of information which people use every day to find relevant information. Lately due to continuous expanding size and complexity of web, retrieving useful information, effectively and efficiently has become a major concern. Various techniques have been proposed both by theoretical researchers and Company researchers. SEO techniques are used by Webmasters for providing better search facility to web crawlers. Web Crawler searches the page for relevant data, those are, keywords, link structure and popularity of page. Based on these finding every search engine create a rank for pages they index. When user send a query, matched or related keywords are searched and results are displayed depending on the scores of websites. User then selects its preferred web-pages. Ranking is based on various parameters, some are known others not. Search engine optimizers work hard to use these parameters and rank their web-pages higher in the result. Every researcher has its own opinion of using SEO techniques. In order to learn how to optimize webpages and their relevancy to search engines page ranking, how search engines use these parameters for gathering information about webpages and what SEO techniques provide better results, we need a detailed study about search engines and their page ranking strategies. To demonstrate the need of effective SEO techniques which are used by search engines, a Meta search engine, *iral*, is presented in this paper. Use of Meta search engines is effective as they cover a larger section of web by sending the query to various search engines. Developing Meta search engines for research such as, using different algorithms for information retrieval [26], [27] to variations in page ranking algorithms [34], [1] has been discussed. Every Meta search engine provides different search result to its users. The effectiveness of a MSE depends on result merging algorithm it provides.

The purpose of iral, Meta search engine is to retrieve data from two major search engines, merge and rank them using SEO parameters and produce the results to users and calculate the relevancy of user query to search results. Every search engine has its own parameters for ranking algorithms to rank webpages. Search engine crawlers use information on webpages such as title name, link structure and others to know the content of webpage. Based on these and other (unknown) parameters page ranking algorithms are formed. The on-page parameters; discussed later, are less effective in generating relevant search results as they are controlled by webmasters. In order to provide a better result experience to users, we combine the SEO parameters with results generated from Meta search engine. By combing parameters on trusted webpages, web spamming will be removed. The ranking algorithm will then generate the results based on SEO parameters. The challenge is to understand the weightage and new SEO techniques for better search results.

Rest of the paper is organized as: in section II gives brief literature review, section III provides experimental approach and results, section IV is about limitation of the research, section V gives conclusion and VI presents future work.

## II. BACKGROUND LITERATURE

### 1) Evolution of Search Engines

Search engines can be defined as the most useful and high profile resources available on the internet. These are powerful tools used to assist the users to find the information in a large pool of data in World Wide Web [10]. According to [24], Explosive growth of World Wide Web resulted in motivation of development of search engines to assist the users for finding desired information from pool of unmanageable data. Search engines present information in real time by using numerous algorithms and are thus better than web directories; which stores information in databases and match query for retrieving results. Some popular search engines are Google, Bing, Alta –Vista and Yahoo.

In the early years when World Wide Web was relatively small finding information on the internet was done by using Web-directories. By 1990's it was observed that human powered categorization model was insufficient for faster and better search in World Wide Web [29]. In Archie; the first ever search engine created; it index files and menu name of FTP by using script based information gatherer [11]. This system supports only name based searches by matching user query to its database. Other models developed later were Veronica [18] which suffers from same problem of indexing. It served the same purpose as Archie while working on plain text file working for Gopher [36]. These early developments of search engines focused mainly on its ability to index web resources and retrieve the best results for users query.

Studies on web search engines were small and publications [38], [31],[ 34] are descriptive in nature. With the introduction of web crawlers, the World Wide Web Worm [28] indexing of documents becomes effective and efficient. The WWWW was used to index the URLs and HTML files; using title string. By 1996, Infoseek and Lycos search tools make huge collection of documents by indexing WWW; allowing users to search using keyword-based queries [15]. With emergence of Google [6] led new standards in search engines. The unique way of ranking retrieved page increases its quality.

Search engines has received good attention lately [4], [8], [12], [17], [18], [21], [32], [35] describes about new algorithms to search dynamic web pages, 3d objects and images.

Fuzzy logic based experimental search engines are described in [2], [24], [37]. Wolfram Alpha is the new generation answer engines which perform computation using information retrieval and NLP (Natural language programming) to answer the query rather than providing list of documents [15].

According to [23], search engines can be classified in three categories; robot based, directory based and Meta search engines; according to information collection and service deliverable. Robot based search engines use software robots to collect and index web sites, download documents in their databases having large indexes. Upon receiving a query they search their database to generate results relevant to query. Examples are Infoseek , Alta Vista and WebCrawler [22]. Directory based search engines organize resources in tree structured directories sorted according to subject area. They collect information by artificial based or by the authors of the websites. Meta search engines are based on multiple individual search engines to extract the data.

### 2) An overview of Meta search engines

A Meta search engine transmits users search query simultaneously to several search engine databases of websites individually and retrieve information from all those search engines queried.

Meta search engines has received lot of attention in academic research, [12], [23], [24], [34] uses meta search engines for generating search results based on their experimental information retrieval techniques. The concept of Meta search engine came from All-in-one Search page in which users select search engines from pool of different search engines. Using intelligent agents to retrieve search results were discussed in [23]. Selecting most promising search engine for information retrieval, SavvySearch [10] was developed in which, query was send parallel to 2-3 search engines but results displayed were not merged. Neural network based intelligent Meta search engines [16], [25]. The ProFusion System [12] supports automatic and manual query dispatching, combing several features of other Meta search engines.

Experimental Meta search engines focus on system architecture such as sorting algorithms [21] and adaptive behavior [12]. Work of [3] uses two models of Meta search engine. Their study was based to enhancing the performance of Meta search engines using two different algorithms, Borda-fuse and Bayes-fuse. ProThes by Braslavaski *et al.,* (2004) revamped the design of Meta search engine by including a thesaurus component.

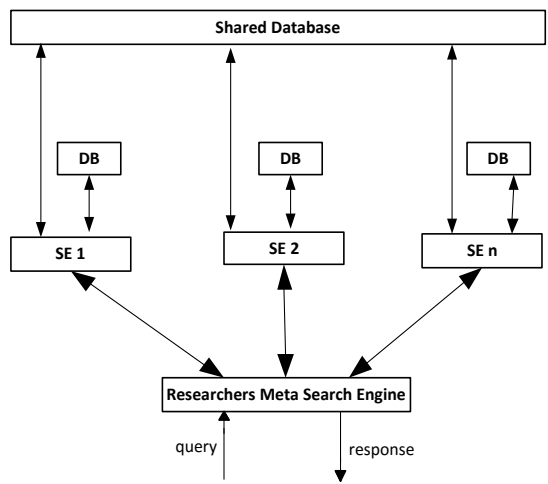

**Figure 1- [3] Architecture of simple Meta search engine**

### 3)   What is Search engine Optimization?

A set of techniques or actions that a web-site can undertake in order to improve its ranking/position in search engine is known as search engine optimization. It essentially involves in providing a higher ranking in search results when a user submit related query to search engines. This ranking is essential for generating higher traffic to the web-sites. Search engines generate their revenue from advertisement and sponsored links. To accommodate advertisers, search engine usually divide their search engine result pages (SERP) into organic and sponsored part. These sponsored links are costly and not every website owner can bid for it. In order to gets a higher page ranking from among millions of websites, owners/developers of websites need to invest time in organic listing. The collection of techniques for improving the organic rank of websites is a process called as search engine optimization.

Search engines are very cautious about SEO techniques (White/ Black hat) and have special guidelines, failing which they remove sites from search results. On February 7, 2006 CNN reported, Google banned BMW's German Website[1] for using doorway techniques, basically showing one page to search engines and different page to users.

The factors for optimizing websites can be categories as: 1. Keyword Analysis (Searching for best suitable keywords) 2.On-page optimization (set of techniques implied to web-pages) 3. Off-page optimization (efforts using social networking and back linking) [1], [20]

#### i)   Keyword Analysis

The quires entered by user in search engines are known as keywords and their combination is collectively known as key phrase. Keywords are important and essential for optimization of any web-site [9]. A thorough research for keywords should be done before choosing a domain name as it accounts for 20% of SEO efforts [9]. It is the first step for optimization and arguably the most important one. An effective keyword is one which has less competition (no. of websites having same keyword) and high volume (number of search for a particular keyword over a period of time) of search. Keywords can either be head terms (one or two words with high search volume) or tail terms (three or more words with less search volume). As an example, '*computer*' is a head term; keyword '*cheap computer for student*' is considered as a tail term. Google suggestion tools are useful for finding the density of any keywords and its global search volume [7].

#### ii)   On-page optimization

These techniques are implied on websites/web-pages for optimization. Search engines 'crawls' the content of websites to know its category or topic area it is relevant to. Advanced tools such as keyword analysis, content analysis and language analysis are used by search engines to categorize a certain web-page [27]. For example a search for cricket should generate those pages which are related to it. Elements which need to be considered optimizing are:

#### (a)   Page title

It tells about the topic of a web-page both to its users and search engines. It is the first element to get crawled by SE's [27]. Google use page title in its result page as a summary. It is written in HTML as,    *<title> content goes here</title>*

#### (b)   Meta Keywords

They are used to define the content of a web-page. They provide bunch of keywords specific to sites content. Most search engines (Google, Yahoo) penalize for abusing its use. They are widely used to provide synonyms to title tags [27]. Syntax:  *<meta name="keywords" content="keyword1, keyword2, keyuword3">*

#### (c)   Meta Description

These tags are used to describe web-pages in a short plain language text. They are often considered as summary of the websites. Ideally they are written in 20-30 words and can be seen in search engines result pages (SREP) below title tags. Syntax: *<meta name="description" content="description goes here">*

#### (d)   Meta Content

It is used to declare character set of website. For example if the document is in UTF-8 punctuation character but been displayed in ASCII or ISO, will not cause any display problem. Syntax: *<meta http-equiv="Content-Type" content="text/html; charset=UTF-8">*

#### (e)   Other optional tags

- Meta Revisit tag

---

[1] www.bmw.de

- Meta robots
- Meta Distribution
- Meta Author
- Meta Language

### (f)     Image Alt attribute

Search engine crawlers fail to read image contents in websites whereas human eyes can interpret the image to its meaning. By providing text description of images, it become easy for crawlers to better understand the meaning of website if it has lot of images embed in it. It is also helpful for image search as the alt text (used to define image) is treated similar as anchor text. Syntax: **

### (g)     Breadcrumb Trial

It is navigational tool used to show user path or hierarchy of website. It is helpful for fast and easy navigation through the website. According to some webmasters it is useful for in-bound linking of web-pages which alternatively gets some points from search engines in ranking. Use of images in breadcrumb helps in earning extra points. Instead of text 'home' hyperlink a 'home' image is more suitable as it can further contain alt text, explained above, and thus treated as anchor text

### (h)     Site Map

Generally sitemaps can be further divided to two types:

(1) HTML page listing which is used for easy navigation for web-users. These are divided into sections and categories depending on the size and pages a website contains.

(2) XML Sitemap- It was introduced by Google for dynamic web-pages. It helps crawlers to crawl webpages developed in Flash or AJAX [27]. It is more precise than normal HTML pages as syntax errors are not tolerated.

Sample coding for XML Sitemap:

```
<?xml version="1.0" encoding="UTF-8"?>
<urlset   xmlns="http://www.sitemaps.org/schemas/sitemap/0.9">   //current
protocol

  <url>    //parent tag for url entry

    <loc>http://www.mysite.com/</loc>   //Page url

    <lastmod>2012-02-21</lastmod> // Last modification date

    <changefreq>weekly</changefreq> //frequency of web-page changing(
hourly,weekly, monthly, never)

    <priority>1.0</priority> // priority of this page in website value0.0-1.0
default-0.5

  </url>

</urlset>
```

### iii)     Off-page optimization

Those strategies which are used outside webpage and is not related to modification on page content. It is more related to page-linking techniques and Social media marketing. They are mostly used to maximize keyword performance of web-sites. Some popular methods are Blog Commenting, posting on forums, bookmarking on social sites, article submission and RSS feed submission. It is a lengthy and continuous process. Anchor text linking is most popular as in this, webmasters use words to hyperlink their websites.  For example, a computer shop website may be linked to a blog discussing computer studies. *<a href="http://mycomputerwebsite.com">Computer Studies</a>*

## III.   PROPOSED APPROACH

### 1)   iral Architecture

In this section architecture of iral is explained.

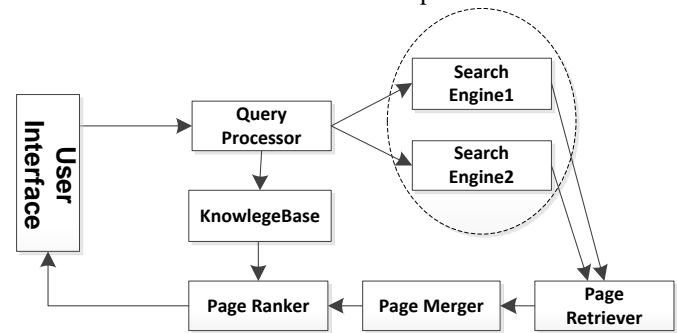

**Figure 2- Proposed iral Architecture**

iral accepts a single query from the user and sends it to multiple search engines. This architecture supports two search engines: Google and Bing.

### a)     User Interface

In present system, User Interface allows user to submit their search queries. The UI deployed here is a search box which supports text based searches. User enters the keyword or combination of keywords for any search query. This query will then be forwarded to query processor by using a Hypertext Pre-processor scripting language.

### b)     Query Processor

Use of query processor is used for explicit notation of user preferences. Query processor is connected to search engines via API (application programming interface) provided by Google and Bing. Other end of query processor is connected to a knowledge base.

### c)     Search Engines

The use of search engines is vital in retrieving results for queries in Meta search engines. The query process send the keyword to search engines, API's maintains all the required setting for the communication with remote search engines. Once the keywords are entered in both search engines, they dispatch the results in HTML format known as Search result records (SRRs).

### d)    Page Retriever

The use of page retriever is to collect all the results from search engines individually. The SRR is dynamically generated web usually enwrapped in HTML tags. It contains useful information such as web-link, domain name, meta-tag, title description and snippet.

### e)    Page Merger

It is an important application which is used for merging the pages form different search engines. The aligned pages from page retriever is send to page merger where it remove any duplicate pages such as same website link from two different search engine. Duplicate webpages will be removed by assigning hash value to every page and sorting them in hash table. Links which produces same hash value will be considered as duplicates and removed.

### f)    Knowledge Base

It is to get user query from query processor and find the synonyms of one work query. As there is no database for the system, knowledge base works by sending the query to dictionary.com via its API. The system then gets the results and send it to page ranker where it search for the related keywords in meta description, snippet and title tags.

### g)    Page Ranker

The purpose of page ranker is to order web-pages according to the ranking algorithm. It use search engine optimization parameters for ranking web-page. The detailed list of parameters and ranking algorithm is described in later section.

## 2)    Ranking Strategy

### a)    PageRank[2]

The concept of PageRank [6] in Google has enormous effects on web optimization techniques and the way webpages are constructed for better ranking.  PageRank lies in assumption that the link in web page is an on topic link and unbiased link to other website. Web spammers have used several methods to manipulate the ranking algorithms, like using; 'link farm', blog spam, paid page linking and massive link exchanges. These assumptions in PageRank highly reduce the effectiveness of result page ranking.

In 1998 Brin and Page found out page rank formula [6] given by:

$$PR(A) = (1-d) + d(\frac{PR(T1)}{C(T1)} + \frac{PR(T2)}{C(T2)}) \qquad (1)$$

In this equation, d is set to 0.85 [6].

*Calculating Page Rank*:

---

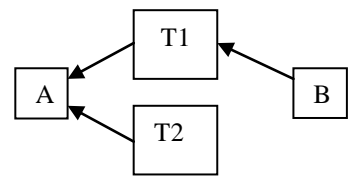

**Figure 3 Linking of webpages**

If  A, T1, T2 and B are webpages and suppose T1 and T2 link page A and B link to Page T1.  C(T) is number of links going out of page. From equation (1) we have,

PR(B)= (1-0.85)+.85 (0) = 0.15
PR(T1)= (1-0.85)+0.85(0.15/1)= 0.27
PR(T2)= (1-0.85)+0.85(0)= 0.15
Hence replacing the value in eq. 1 we get,
PR(A)= (1-0.85)+0.85(0.27+0.15)= 0.5133

Proof,
As stated in [6] sum of all the links is equal to 1
PR(A)+ PR(T1)+ PR(T2)+ PR(B)= 0.5133+0.27+0.15+0.15
= 1

From recent studies and news it's obvious that search engines keep their algorithms safe and update them frequently. Algorithms are now more sophisticated and improved from the past.

### b)    Advanced iral Ranking Method

The ranking parameter of page ranking algorithm is modified according to need of this experiment. Using combinations of ranking metrics, algorithms explained above and combing them with other variables a new parameter for setting the page rank of webpages can be derived. The new parameters for ranking have been assigned using following parameters:

**Table 1: Page Ranking parameters**

| SEO Parameters | Description |
|---|---|
| Title Tag | If query is a term in webpage title |
| Meta Description | If query term is present in description |
| Meta Keyword | If query matches keyword |
| Snippet | Number of times word appears in summary |
| Meta expires | Recent pages to be given higher rank |
| Meta content | help to show compatible pages to users (UTF-8, ASCII or ISO) |
| Image attribute | use of Image alt attribute helps in higher weightage as it provides description to images |
| Sitemap | use of sitemap increase visibility for dynamic web-pages |
| Links present | number of in-bound links to the page |

For every parameter same weightage is assigned. This is to understand the ranking weightage of different search engines.

---

[2] The name PageRank is a trademark of Google.

Every important[3] search engine optimizing (SEO) factors are considered and assigned the weight.

The ranking is calculated as below:

$$PageRank(P) = \sum v_i(P) * w_i(v_i) \quad (2)$$

In equation (2): $v_i$ is the value of parameter and $w_i$ is the weight of the parameter $v_i$

### 3) Experiment Setup- iral Meta search engine

To see the effects of web-optimization techniques in search results and their relevance to the user query, the following online experiment was performed. The first batch of experiments was performed using 2 types of keywords: head keywords (they consist of single word) and tail keywords (they are longer in length 3 or more words). This batch included 300 students randomly selected. They were asked to use below keywords and rate the search engines in a scale of 5, where 1 is the least preferred and 5 is the most.

Keywords used: 'alcoholism' and 'local computer shop'

| Search Engine | Mean Precision |
|---|---|
| Google | 0.44 |
| Bing | 0.31 |
| iral Meta search | 0.48 |

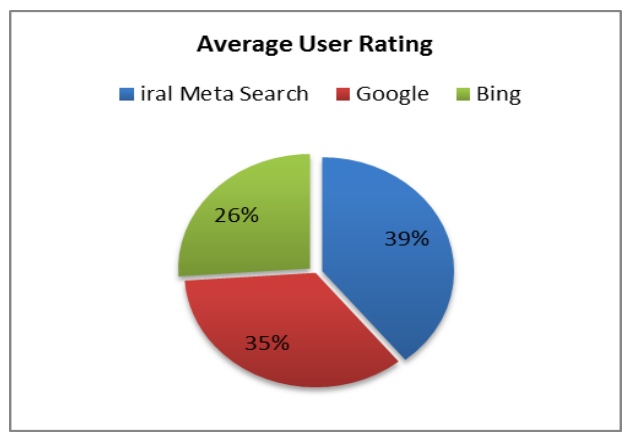

### IV. LIMITATIONS

This paper offers quantitative methods for obtaining the results. There are many factors which affect the ranking of any webpage. These factors are kept secret by search engines to keep secure their ranking algorithms. General SEO techniques were used for experiments but it was hard to draw any general conclusion from retrieved results due to limitation of queries used and people engaged.

---

[3] Only white hat techniques are used

### V. CONCLUSION

We presented an approach for using SEO technique for retrieving relevant data from search engines. The iral ranking algorithm uses search engine optimization factors to assign heuristic page rank to search engine result pages and display them to the users. Experimental results have proved that iral Meta search engine with average precision of 0.48 has outperformed Google (0.44) and Bing (0.31) search engines. Initial results were promising and findings revealed that search engine optimization is useful in categorizing web data effectively and ranking based on these parameters can achieve results more beneficial for user query thus increasing user satisfaction.

Advanced iral Meta search ranking algorithm has proved to be effective and efficient for head term keywords. The iral Meta search engine is online for users and there is interesting possibility for learning lifetime. The system will improve over the time and interactive feedbacks from user will help in future changes.

Interestingly, it was observed that high level of web-page optimization is useful when search algorithms of search engines are less accurate. Results also provides important recommendation to webmasters that optimization to certain level helps website ranking and indexing but using black hat SEO techniques may cause webpage been banned from search engines.

Future work involves, incorporating more search engines in the study, providing more queries and have a larger user base. Parameters for page ranking change frequently; an intense study is needed for throwing light to this ever changing topic.

### VI. REFERENCES

[1] Albert Bifet and Carlos Castillo, "An Analysis of Factors Used in Search Engine Ranking", In Proceedings of the 14th International World Wide Web Conference (WWW2005), First International Workshop on Adversarial Information Retrieval on the Web (AIRWeb'05), 2005

[2] Aquin Motta, M. d. A. a. E. (2011). "Watson, more than a Semantic Web search engine." Semantic Web **2**(1): 55-63

[3] Aslam, J. A. and M. Montague (2001). Models for metasearch. Proceedings of the 24th annual international ACM SIGIR conference on Research and development in information retrieval. New Orleans, Louisiana, United States, ACM**:** 276-284.

[4] Bajracharya, S., T. Ngo, et al. (2006). Sourcerer: a search engine for open source code supporting structure-based search. Companion to the 21st ACM SIGPLAN symposium on Object-oriented programming systems, languages, and applications. Portland, Oregon, USA, ACM**:** 681-682.

[5] Braslavski, P., Alshanski, G., & Shishkin, A. (2004). ProThes: Thesaurus based meta-search engine for a specific application domain. Proceedings of the 13th International WorldWideWeb Conference, 222–223.

[6] Brin S. and Page L. The anatomy of a large-scale hypertextual search engine. In Proc. 7th Int. World Wide Web Conference, 1998, pp. 107–117.

[7] Clifton, Brian (2012)http://www.amazon.com/gp/product/1118168445/ref=as_li_tf_tl?ie=UTF8&tag=advawebmetr-20&linkCode=as2&camp=1789&creative=9325&creativeASIN=1118168445#reader_B007RSUSZ0

[8] Cohen, S., J. Mamou, et al. (2003). XSEarch: a semantic search engine for XML. Proceedings of the 29th international conference on Very large data bases - Volume 29. Berlin, Germany, VLDB Endowment: 45-56.

[9] David Viney, http://www.amazon.com/Get-Top-Google-Techniques-Rankings/dp/1857885023#reader_1857885023

[10] Dreilinger, D. and A. E. Howe (1997). "Experiences with selecting search engines using metasearch." ACM Trans. Inf. Syst. 15(3): 195-222.

[11] Emtage, A. (1993). Archie Intelligent Information Retrieval: The Case of Astronomy and Related Space Sciences. A. Heck and F. Murtagh, Springer Netherlands. 182: 103-111.

[12] Fan, Y. and Gauch, S. (1999), "Adaptive agents for information gathering from multiple, distributed information sources", available at:www.ittc.ukans.edu/.sgauch/papers/AAAI99.doc

[13] Ferragina, P. and A. Gulli (2008). "A personalized search engine based on Web-snippet hierarchical clustering." Software: Practice and Experience 38(2): 189-225.

[14] Funkhouser, T., P. Min, et al. (2003). "A search engine for 3D models." ACM Trans. Graph. 22(1): 83-105.

[15] Gauch "ProFusion: Intelligent Fusion from Multiple, Distributed Search Engines," Susan Gauch, Guijun Wang and Mario Gomez, Journal of Universal Computer Science, Vol. 2, No. 9, September 1997, 637-649.

[16] Gray, J. (2012). "Interview with Dr Stephen Wolfram." Linux J. 2012(214): 5.

[17] Harley Hahn and Rick Stout. The Internet Complete Reference. Osborne McGraw-Hill, Berkeley, California, 1994.

[18] Hogan, A., A. Harth, et al. (2011). "Searching and browsing Linked Data with SWSE: The Semantic Web Search Engine." Web Semantics: Science, Services and Agents on the World Wide Web 9(4): 365-401.

[19] Horowitz, D. and S. D. Kamvar (2010). The anatomy of a large-scale social search engine. Proceedings of the 19th international conference on World wide web. Raleigh, North Carolina, USA, ACM: 431-440.

[20] Huang, L., Hemmje, M. and Neuhold, E.J. (2000), "ADMIRE: an adaptive data model for meta searchengines", available at: http://www9.org/w9cdrom/165/165.html

[21] Ledford, J. (2009). SEO: Search Engine Optimization Bible, Wiley Publishing.

[22] Lei, Y., V. Uren, et al. (2006). SemSearch: A Search Engine for the Semantic Web Managing Knowledge in a World of Networks. S. Staab and V. Svátek, Springer Berlin / Heidelberg. 4248: 238-245.

[23] Leighton, H.V. & Srivastava, J. (1997). Precision among World Wide Web search services (search engines): Alta Vista, Excite, HotBot, Infoseek, Lycos [online]. Available at http://www.winona.msus.edu/library/webind2/webind2.htm

[24] Lin, G., J. Tang, et al. (2011). Studies and evaluation on meta search engines. Computer Research and Development (ICCRD), 2011 3rd International Conference on.

[25] Liu, H., H. Lieberman, et al. (2006). GOOSE: A Goal-Oriented Search Engine with Commonsense Adaptive Hypermedia and Adaptive Web-Based Systems. P. De Bra, P. Brusilovsky and R. Conejo, Springer Berlin / Heidelberg. 2347: 253-263.

[26] M. Beigi, A. Benitez, S.-F. Chang, Metaseek: A content-based meta search engine for images, Proc. SPIE Storage and Retrieval for Image and Video Databases, San Jose, CA, 1998, http://www.ctr.columbia.edu/metaseek

[27] Mahabhashyam, S. M. and Singitham, P.; Tadpole: A Meta search engine Evaluation of Meta Search ranking strategies; University of Stanford; 2004.

[28] Manoj, M. and Elisabeth Jacob. Information retrieval on internet using meta-search engines: A review. Journal of Scientific & Industrial Research, 67:739{746, 2008.

[29] Odom, Sean and Allison, Lynell (2012) http://www.amazon.co.uk/Seo-2012-Sean-Odom/dp/0984860002/ref=sr_1_1?ie=UTF8&qid=1334861867&sr=8-1#reader_0984860002

[30] Oliver A. McBryan. GENVL and WWWW: Tools for Taming the Web. First International Conference on the World Wide Web. CERN, Geneva (Switzerland), May 25-26-27 1994.

[31] P. Metaxas. On the evolution of search engine rankings. In Proceedings of the 5th WEBIST Conference, Lisbon, Portugal, March 2009.

[32] Poaolo Boldi, Massimo Santini, Sebatiano Vigna, "Pagerank as a function of damping factor" DSI Università degli Studi di Milano.

[33] Shirky, Clay. (October, 1995). Finding needles in haystacks. Netguide, 87-90.

[34] Strohman, T., Metzler, D. and Turtle, H., and Croft, B. (2004). Indri: A Language Model-based Search Engine for Complex Queries. In Proceedings of the International conference on Intelligence Analysis.

[35] T. Joachims. Optimizing search engines using clickthrough data. In Proceedings of the ACM 2002.

[36] Taubes, Gary. (September 8, 1995). Indexing the Internet. Science, 269, 1354-1356.

[37] Theobald, A. and G. Weikum (2002). The Index-Based XXL Search Engine for Querying XML Data with Relevance Ranking Advances in Database Technology — EDBT 2002. C. Jensen, S. Šaltenis, K. Jefferyet al, Springer Berlin / Heidelberg. 2287: 311-340.

[38] Weiss, R., B. V\, et al. (1996). HyPursuit: a hierarchical network search engine that exploits content-link hypertext clustering. Proceedings of the the seventh ACM conference on Hypertext. Bethesda, Maryland, United States, ACM: 180-193.

[39] Widyantoro, D. H. and J. Yen (2001). A fuzzy ontology-based abstract search engine and its user studies. Fuzzy Systems, 2001. The 10th IEEE International Conference on.

[40] Wildstrom, Stephen H. (September 11, 1995). Feeling your web around the Web. Business Week, 22.